\documentclass[12pt,fleqn]{article}
\usepackage{amsfonts,amssymb,latexsym,cite}
\usepackage{epsf,graphicx}

\def\noi{\noindent}

\newcommand{\Title}[1]{\section*{\Large\bf #1}}

\def\Aunames#1{\noi{\large\bf #1}}
\def\auth#1{${}^{#1}$}
\def\Addresses#1{\medskip\noi \protect
    \begin{description}\itemsep -3pt {\it #1} \end{description}}
\def\addr#1#2{\item[${}^{#1}$]{\it #2}}

\newcommand{\Abstract}[1]{\vskip 2mm \begin{center}
        \parbox{16.4cm}{\small\noi #1} \end{center}\medskip}

\def\email#1#2{\footnotetext[#1]{e-mail: #2}\addtocounter{footnote}{1}}

\def\nqq{\hspace*{-2em}}
\def\nhq{\hspace*{-0.5em}}

\def\cm{\hspace*{1cm}}

\def\Jl#1#2{#1 {\bf #2},\ }

\def\ApJ#1 {\Jl{Astroph. J.}{#1}}
\def\CQG#1 {\Jl{Class. Quantum Grav.}{#1}}
\def\DAN#1 {\Jl{Dokl. AN SSSR}{#1}}
\def\GC#1 {\Jl{Grav. Cosmol.}{#1}}
\def\GRG#1 {\Jl{Gen. Rel. Grav.}{#1}}
\def\JETF#1 {\Jl{Zh. Eksp. Teor. Fiz.}{#1}}
\def\JETP#1 {\Jl{Sov. Phys. JETP}{#1}}
\def\JHEP#1 {\Jl{JHEP}{#1}}
\def\JMP#1 {\Jl{J. Math. Phys.}{#1}}
\def\NPB#1 {\Jl{Nucl. Phys. B}{#1}}
\def\NP#1 {\Jl{Nucl. Phys.}{#1}}
\def\PLA#1 {\Jl{Phys. Lett. A}{#1}}
\def\PLB#1 {\Jl{Phys. Lett. B}{#1}}
\def\PRD#1 {\Jl{Phys. Rev. D}{#1}}
\def\PRL#1 {\Jl{Phys. Rev. Lett.}{#1}}

\def\al{&\nhq}
\def\lal{&&\nqq {}}

\def\eqs{Eqs.\,}
\def\beq{\begin{equation}}
\def\eeq{\end{equation}}
\def\bear{\begin{eqnarray}}
\def\bearr{\begin{eqnarray} \lal}
\def\ear{\end{eqnarray}}
\def\earn{\nonumber \end{eqnarray}}
\def\nn{\nonumber\\ {}}

\def\nnn{\nonumber\\ \lal }
\def\nnnv{\nonumber\\[5pt] \lal }

\def\eql{\al =\al}

\def\dst{\displaystyle}
\def\tst{\textstyle}
\def\fracd#1#2{{\dst\frac{#1}{#2}}}
\def\fract#1#2{{\tst\frac{#1}{#2}}}
\def\Half{{\fracd{1}{2}}}
\def\half{{\fract{1}{2}}}

\def\e{{\,\rm e}}
\def\d{\partial}

\def\sign{\mathop{\rm sign}\nolimits}
\def\diag{\mathop{\rm diag}\nolimits}

\def\const{{\rm const}}

\def\mn{_{\mu\nu}}
\def\MN{^{\mu\nu}}
\def\mN{_\mu^\nu}

\def\wh{wormhole}
\def\whs{wormholes}

\def\ssph{static, spherically symmetric}
\def\asflat{asymptotically flat}

\def\oR{\overline{R}}
\def\og{\overline{g}}

\def\M{{\mathbb M}}

\def\cT{{\cal T}}
\def\ME {\mbox{$\M_{\rm E}$}}
\def\MJ {\mbox{$\M_{\rm J}$}}
\def\fin{{\rm fin}}

\begin{document}

\Title{Once again on thin-shell wormholes in scalar-tensor gravity}

\Aunames{Kirill A. Bronnikov\auth{a,1} and Alexei A. Starobinsky\auth{b,c,2}}

\Addresses{
\addr a {\small Center of Gravitation and Fundamental Metrology,
   VNIIMS, 46 Ozyornaya St., Moscow 117361, Russia; \\
   Institute of Gravitation and Cosmology,
        PFUR, 6 Miklukho-Maklaya St., Moscow 117198, Russia  }
\addr b {\small Landau Institute for Theoretical Physics of RAS,
        Moscow 119334, Russia}
\addr c {\small RESCEU, Graduate School of Science, The University of
                Tokyo, Tokyo 113-0033, Japan}
           }

\bigskip

\Abstract{It is proved that all thin-shell wormholes built from two identical
  regions of vacuum static, spherically symmetric space-times have a negative
  shell surface energy density in any scalar-tensor theory of gravity with a
  non-ghost massless scalar field and a non-ghost graviton.}

\email 1 {kb20@yandex.ru}
\email 2 {alstar@landau.ac.ru}

\bigskip \bigskip

  It has been recently proved in a general form \cite{we07} that no wormholes
  can be formed in any scalar-tensor theory (STT) of gravity in which the
  non-minimal coupling function $f(\Phi)$ is everywhere positive and the
  scalar field $\Phi$ itself is not a ghost if matter sources of gravity
  respect the Null Energy Condition (NEC).

  In order to construct any viable and stable \wh\ solution, attempts have
  been recently made to find such a solution in STT of gravity using thin 
  shells instead of extended matter sources, and it has been claimed that at
  least in some cases (namely, in the Brans-Dicke STT for a particular range
  of values of the coupling constant $\omega$) the shell at the \wh\ throat
  may satisfy the weak and null energy conditions \cite{eir08}.
  In this short note we will show explicitly that, in {\it any\/} STT with a
  massless non-ghost scalar field, in {\it all\/} thin-shell \whs\ built from
  two identical regions of vacuum \ssph\ space-times, the shell has negative
  surface energy density (note that we here do not consider numerous toy \wh\
  models which do not represent solutions of any initially fixed equations of
  some metric theory of gravity).

  Let us begin with presenting the nonzero components of the Einstein tensor
  $G\mN = R\mN - \half \delta\mN R$ for a general \ssph\ space-time with the
  metric\footnote
    {Our conventions are: the metric signature $(+{}-{}-{}-)$; the
    curvature tensor
    $R^{\sigma}{}_{\mu\rho\nu} = \d_\nu\Gamma^{\sigma}_{\mu\rho}-\ldots,\
    R\mn = R^{\sigma}{}_{\mu\sigma\nu}$, so that the Ricci scalar
    $R > 0$ for de Sitter space-time and the matter-dominated
    cosmological epoch; the system of units $8\pi G = c = 1$.}
\beq
    ds^2 = \e^{2\gamma(u)}dt^2                                    \label{ds}
                - \e^{2\alpha(u)} du^2 - \e^{2\beta(u)} d\Omega^2,
\eeq
  where $d\Omega^2 = d\theta^2 + \sin^2\theta d\zeta^2$ is the metric on a
  unit sphere and $u$ is an arbitrarily chosen radial coordinate. We
  have (the prime denotes $d/du$)
\bear                                                            \label{Gmn}
    G^0_0 \eql \e^{-2\alpha}
            (2\beta''+  3\beta'{}^2-2\alpha'\beta') - \e^{-2\beta},
\nn
    G^1_1 \eql \e^{-2\alpha}
            (\beta'{}^2 + 2\beta'\gamma'] - \e^{-2\beta},
\nn
    G^2_2 = G^3_3 \eql \e^{-2\alpha}
            [\gamma'' + \gamma'{}^2 + \beta'' + \beta'{}^2
            + \beta'\gamma' - \alpha'(\beta'+\gamma')].
\ear

  Now, consider Fisher's well-known solution \cite{fish} to the Einstein
  --- massless scalar equations which can be written in the form
\bear                                                             \label{F1}
    ds^2 \eql P^a dt^2 - P^{-a} dr^2 - P^{1-a}r^2 d\Omega^2,
\\                                                                \label{F2}
    \psi \eql - \frac{C}{2k} \ln P(r),
\\                                                                \label{F3}
    a^2 \eql 1 - \frac{C^2}{2k^2},\cm
            P = P(r) := 1 - \frac{2k}{r},
\ear
  where $\psi$ is the scalar field, $a$, $C$ and $k$ are integration
  constants related as given in (\ref{F3}). The metric (\ref{F1}) is
  \asflat, with the Schwarzschild mass $m$ equal to $ak$; note that
  $a^2 \leq 1$, and in case $a=1$, $C=0$ the Schwarzschild solution is
  restored.

  The solution (\ref{F1})--(\ref{F3}) with $C\ne 0$ has a naked singularity
  at $r=2k$, situated at the centre of symmetry since there $g_{22}=0$ (the
  coordinate spheres shrink to a point). However, following \cite{eir08} and
  some other papers, one can easily obtain a traversable \wh\ geometry using
  the cut-and-paste trick \cite{V89}: one takes two copies of the region
  $r \geq r_0 > 2k$ of the space-time (\ref{F1}) and identifies the spheres
  $r=r_0$ in them. This procedure is formally described as putting
\beq                                                              \label{u}
      r = r_0 + |u|
\eeq
  in (\ref{F1})--(\ref{F3}),
  where $u$ is a new radial coordinate, and assigning $u\leq 0$ to one copy
  of the region $r\geq r_0$ and $u\geq 0$ to the other. Then the derivatives
  of $g\mn$ are discontinuous at $u=0$, and this may be ascribed to appearance
  of a thin shell with certain energy density and tension. The latter can be
  found by substituting the solution (\ref{F1})--(\ref{F3}) rewritten in
  terms of $u$ according to (\ref{u}) into the Einstein equations
\beq
     G\mN = - S\mN (\psi) - \cT\mN,                         \label{EE}
\eeq
  where $S\mN(\psi) = \psi^{,\nu}\psi_{,\mu} -\half \delta\mN (\d\psi)^2$
  is the stress-energy tensor (SET) of the field $\psi$ while $\cT\mN$ is the
  shell SET proportional to Dirac's delta function, $\delta(u)$. Outside the
  shell, \eqs (\ref{EE}) are manifestly satisfied by our solution, and the
  task is to find
\beq
    \cT\mN = \delta(u) \diag (\sigma,\ 0,\ -p_\bot,\ -p_\bot),
\eeq
  where $\sigma$ is the surface density and $p_\bot$ the surface pressure;
  the radial pressure should evidently vanish because the shell is
  perpendicular to the radial direction.

  Nonzero contributions to $\sigma$ and $p_\bot$ appear only due to
  $d^2 r/du^2 = 2\delta(u)$. Therefore, to find them, in the expressions
  (\ref{Gmn}) it is sufficient to take into account only terms with
  second-order derivatives since all other terms are finite at $u=0$ (to be
  denoted by the symbol ``$\fin$''). In particular, $T\mN(\psi) = \fin$ and
  $G^1_1 = \fin$ because they contain only first-order derivatives.

  A straightforward calculation then gives
\bear
    \cT^0_0 \eql -\frac{4\delta(u)}{r_0^2} P_0^{a-1} [r_0 - (1+a)k] + \fin,
          \cm \cT^1_1 = \fin,
\nn                                                             \label{TE}
    \cT^2_2 = \cT^3_3
        \eql -\frac{2\delta(u)}{r_0^2} P_0^{a-1} (r_0 - k) + \fin,
\ear
  where $P_0 = P(r_0) = 1-2k/r_0$. Since $r_0 > 2k$ and $|a| <1$, the surface
  density $\sigma$ is negative.

  We have thus calculated the shell characteristics directly from the
  Einstein equations without invoking the well-known Israel formalism
  for thin shells; our method is similar to what was done in \cite{SS77}
  in the case of conical singularities (infinitely thin cosmic strings).
  The Israel formalism would be quite necessary if we wished to study the
  shell dynamics, but in a purely static description our direct method is
  simpler and more transparent.

  Now, we can recall that the Einstein-scalar equations, whose solution is
  given by (\ref{F1})--(\ref{F3}), can be regarded as the Einstein-frame
  equations of an arbitrary STT with the Jordan-frame Lagrangian
  (written in the Brans-Dicke parametrization for a space-time manifold \MJ\
  with the metric $g\mn$)
\beq                                                             \label{LJ}
     L_{\rm J}= \Half \left[\phi R + \frac{\omega(\phi)}{\phi}
            g^{\mu\nu}\phi_{,\mu} \phi_{,\nu}- 2U(\phi)\right] + L_m,
\eeq
  where $R$ is the Ricci scalar, $L_m$ is the Lagrangian of nongravitational
  matter, $\omega(\phi)$ and $U(\phi)$ are arbitrary functions. In the
  general case, transition to the Einstein frame, defined as a manifold
  \ME\ with the metric
\beq
    \og\mn = |\phi| g\mn,                                  \label{conf}
\eeq
  results in the Lagrangian
\beq                                                        \label{LE}
     L_{\rm E} = \Half (\sign \phi) \biggl[\oR
               + [\sign(\omega+3/2)] \og\MN \psi_{\mu}\psi_{,\nu}\biggr]
                    - V(\psi) + L_{\rm mE},
\eeq
  where bars mark quantities obtained from or with $\og\mn$, indices are
  raised and lowered with $\og\mn$ and
\beq                                \label{trans}
        \frac{d\psi}{d\phi} = \frac{\sqrt{|\omega + 3/2|}}{|\phi|},
    \cm
    V(\psi) = \phi^{-2} U(\phi).
    \cm
    L_{\rm mE} = \phi^{-2} L_m.
\eeq
  The above relations describing a thin-shell \wh\ represent a solution to
  the field equations corresponding to the Lagrangian (\ref{LE}), where the
  matter Lagrangian $L_{\rm mE}$ leads to the SET (\ref{TE}), the potential
  $V(\psi) \equiv 0$ (the scalar field is massless), and the following sign
  conditions hold:  $\phi > 0$ (which means that the graviton is not a ghost)
  and $\omega + 3/2 > 0$ (the $\psi$ field has a normal sign of kinetic
  energy, and, equivalently, the $\phi$ field is not a ghost).

  Suppose a particular STT is chosen, satisfying the above sign conditions,
  with a certain function $\omega(\phi)$ and $U(\phi)\equiv 0$. Then,
  according to (\ref{conf}) and (\ref{trans}), the metric
\beq                                                             \label{dsJ}
   ds_J^2 = \frac{1}{\phi}
        \Big[ P^a dt^2 - P^{-a} dr^2 - P^{1-a}r^2 d\Omega^2 \Big],
\eeq
  with the notations (\ref{F3}) and the scalar field $\phi$ related to
  $\psi$ as given in (\ref{trans}), satisfy the field equations due to
  (\ref{LJ}) with $L_m = 0$ and represent a scalar-vacuum solution of the
  theory (\ref{LJ}), discussed in a general form in \cite{br73}. Moreover, the
  result of the cut-and-paste procedure described above, applied to
  (\ref{dsJ}) with the substitution (\ref{u}), is, in general, a \wh\
  configuration with a thin shell at the throat $r=r_0$ whose SET $T\mN$ has
  the form $T\mN = \phi^2 \cT\mN$ with $\cT\mN$ given by (\ref{TE}). This
  means that such \whs\ are supported by thin shells with negative energy
  density in {\it any\/} STT. This is a direct consequence of the relations
  between the Jordan and Einstein frames.

  For better clarity, let us confirm this conclusion by a direct calculation.
  The metric field equations following from (\ref{LJ}) read
\bear                                                             \label{JE}
       (G\mN + \nabla_\mu \nabla^\nu - \delta\mN \Box)\phi
     + \frac{\omega(\phi)}{\phi} S\mN(\phi) + \delta\mN U(\phi)
        = - T\mN,
\ear
  where $\Box = \nabla_\alpha \nabla^\alpha$ is the d'Alembert operator;
  $S\mN (\phi) = \phi^{,\nu}\phi_{,\mu} -\half \delta\mN (\d\phi)^2$;
  in the case under consideration, $U\equiv 0$;
  and $T\mn = (2/\sqrt{-g})(\delta L_m \sqrt{-g}/\delta g\MN)$
  is the SET corresponding to $L_m$ (all operations being performed with the
  metric $g\mn$ involved in (\ref{LJ})).

  Substituting the metric (\ref{dsJ}) with (\ref{u}) into (\ref{JE}), we are
  again interested only in terms proportional to $\delta(u)$. Therefore, we
  can write
\bearr
        G^0_0 = 2\e^{-2\alpha} \beta'' + \fin, \cm
        G^1_1 = \fin, \cm
        G^2_2 = \e^{-2\alpha} (\beta'' + \gamma'') + \fin,
\nnn
    \beta'' = \frac{2\delta(u)}{r^2 P}[r-k(1+a)]
            - \Half \frac{\phi''}{\phi} + \fin,
\nnn
    \gamma'' = \frac{2ak\,\delta(u)}{r^2 P}
            - \Half \frac{\phi''}{\phi} + \fin,
\nnn
    (\nabla^0\nabla_0 - \Box)\phi =
    (\nabla^2\nabla_2 - \Box)\phi + \fin = \e^{-2\alpha}\phi'' + \fin,
\nnnv
    (\nabla^1\nabla_1 - \Box)\phi = \fin,   \cm
    S\mN (\phi) = \fin,                                     \label{fs}
\ear
  where the metric (\ref{dsJ}) is identified with (\ref{ds}), so that,
  e.g., $\e^{-2\alpha} = \phi P^a$; as before, $P=P(r)$ and the prime means
  $d/du$. As a result, we obtain
\bear
     T^0_0 \eql -\frac{4\phi^2\delta(u)}{r_0^2} P_0^{a-1} [r_0 - (1+a)k]
        + \fin,   \cm T^1_1 = \fin,
\nn                                                             \label{TJ}
     T^2_2 = T^3_3
        \eql -\frac{2\phi^2\delta(u)}{r_0^2} P_0^{a-1} (r_0 - k) + \fin,
\ear
  so that $T\mN = \phi^2 \cT\mN$, as required. It is of interest that
  this calculation did not involve the relation between $\phi$ and $\psi$
  from (\ref{trans}); it was only assumed that $\phi(r)$ is finite and
  smooth, so that only $\phi''$ could lead to expressions with $\delta(u)$.

  Let us use the Brans-Dicke theory, with $\omega = \const > -3/2$, as a
  particular example. In this case, as follows from (\ref{trans}), we can
  put
\beq
    \phi = \exp[\psi/\sqrt{\omega+3/2}] = P^\xi,\cm
    \xi := -\frac{C}{2k\sqrt{\omega+3/2}},
\eeq
  where $\xi$ is an effective scalar charge. The relation between constants
  in (\ref{F3}) is converted to
\beq
    a^2 + (2\omega + 3) \xi^2 =1.
\eeq
  Then in (\ref{dsJ}) and (\ref{fs}) we have
\[
    \e^{2\gamma} = P^{a-\xi}, \cm   \e^{-2\alpha} = P^{a+\xi}, \cm
    \e^{2\beta} = rP^{1-a-\xi},
\]
  and a direct calculation confirms the result (\ref{TJ}).\footnote
    {Though we are using the same coordinate $r$ in the same Brans-Dicke
    scalar-vacuum solution as in \cite{eir08}, our notations
    for the constants are different: their constants $\eta,\ A,\ B$ are
    equal to our $k,\ a-\xi,\ -(a+\xi)$, respectively.}

  Inclusion of radial electromagnetic fields would make the solutions
  slightly more complex and diverse \cite{br73} but the main result
  would be the same: in a ghost-free STT, thin-shell \whs\ can be obtained
  with negative surface shell densities only, violating the weak energy
  condition.

  For comparison, it can be recalled that if we admit $\omega < -3/2$ in the
  Lagrangian (\ref{LJ}), which makes the $\phi$ field a ghost, then \whs\
  exist already among vacuum and electrovacuum solutions of such
  theories \cite{br73} without any shells. It has been shown, however, that
  at least some vacuum solutions of this class are unstable 
  \cite{instab1, instab2} (see also \cite{DHNS08}).

\medskip
  AS acknowledges RESCEU hospitality as a visiting professor. He was
  also partially supported by the grant RFBR 08-02-00923 and by the
  Scientific Programme ``Astronomy'' of the Russian Academy of Sciences.
  KB acknowledges hospitality at Institut de Ci\`encies de l'Espai
  (Barcelona), where part of the work was done, and partial support from
  NPK MU grant of Peoples' Friendship University of Russia.

\end{document}